\documentclass[10pt]{article}
\usepackage[latin1]{inputenc}
\newcommand\beq{\begin{equation}}
\newcommand\eeq{\end{equation}}
\newcommand\bea{\begin{eqnarray}}
\newcommand\eea{\end{eqnarray}}

\usepackage{amsmath,amsfonts,amsthm,graphicx,latexsym,eufrak,bbm,mathrsfs,parskip,latexsym,amssymb}
\usepackage{bbold}
\begin{large}
\author{Kaushlendra Kumar\footnote{School of Biochemical Engineering,Indian Institute of Technology(BHU), Varanasi 221005, India}~ Shivraj Prajapat$^{\ddagger}$\footnote{Indian Institute of Science Education and Research(IISER), Pashan Road, Pune 411008, India ,Email: sapotra@students.iiserpune.ac.in}\\ ~  Biswajit Chakraborty\footnote{S.N.Bose National Centre For Basic Sciences, Salt Lake, Kolkata 700098, India, Email: biswajit@bose.res.in}
}
\end{large}
\begin{LARGE}
\title{On the role of Schwinger's SU(2) generators for Simple Harmonic Oscillator in 2D Moyal plane}
\end{LARGE}

\begin{document}
\maketitle

\begin{abstract}
The Hilbert-Schmidt operator formulation of non-commutative quantum mechanics in 2D Moyal plane is shown to allow one to construct Schwinger's SU(2) generators. Using this the SU(2) symmetry aspect of both commutative and non-commutative harmonic oscillator are studied and compared. Particularly, in the non-commutative case we demonstrate the existence of a critical point in the parameter space of mass($\mu$) and angular frequency($\omega$) where there is a manifest SU(2) symmetry for a unphysical harmonic oscillator  Hamiltonian built out of commuting (unphysical yet covariantly transforming under SU(2)) position like observable. The existence of this critical point is shown to be a novel aspect in non-commutative harmonic oscillator, which is exploited to obtain the spectrum and the observable mass ($\mu$) and angular frequency ($\omega$) parameters of the physical oscillator-which is generically different from the bare parameters occurring in the Hamiltonian. Finally, we show that a Zeeman term in the 
Hamiltonian of non-commutative physical harmonic oscillator, is solely responsible for both SU(2) and time reversal symmetry breaking.
\end{abstract}

\section{Introduction}
It has been realized some time back that the localization of an event in space-time with arbitrary  accuracy is not possible \cite{b1}. This idea was also corroborated by considering certain low energy implication in string theory \cite{b2}. Ever since then there have been an upsurge of literature involving formulation of quantum mechanics and quantum field theory in these kind of spaces where space-time coordinates are upgraded to the level of operators satisfying non-vanishing commutator algebra. The simplest of such quantized space-time is of Moyal type and the corresponding commutator algebra is given by the following $[\hat{x}_{\mu},\hat{x}_{\nu}] = i\theta_{\mu\nu}.$ Generally, quantum mechanics and quantum field theory in such space-time are investigated by demoting operator valued coordinate to c-numbers, at the price of replacing point-wise multiplication of fields by star multiplication \cite{b3}. However the star operation is not unique and can give rise to inequivalent physical consequences \cite{
b4,b5}. It is therefore desirable  to start with the formulation of quantum mechanics itself at completely 
operatorial level, so that one can formulate second quantization and eventually construct quantum field theory at a completely operatorial level. Indeed, such an attempt was made in \cite{a6},albeit in a nonrelativistic framework using the so called Hilbert-Schmidt operator which was initiated in \cite{b6,b7} , with the operator valued space coordinates were taken to correspond to the Moyal-plane satisfying \\
\begin{equation}
[\hat{x}_{i},\hat{x}_{j}] = i\theta_{ij}=i\theta\epsilon_{ij}~~~(i,j=1,2) \label{i1}
\end{equation}
and time was taken to be c-number variable. Clearly, the representation of this coordinate algebra (\ref{i1}) is furnished by a Hilbert space  isomorphic to the quantum Hilbert space of 1-d harmonic oscillator and we refer to this space as classical Hilbert space($\mathcal{H}_c$), which is nothing but boson Fock space
\begin{equation}
\mathcal{H}_c = \text{span} \lbrace \mid n\rangle =\frac{1}{\sqrt{n!}}(\hat{b}^\dagger)^n|0\rangle\rbrace _{n=0}^{\infty}\label{i2}
\end{equation}
where\\
\begin{equation}
\hat{b}= \frac{1}{\sqrt{2\theta}}\left( \hat{x}_1 + i \hat{x}_2\right) \label{i3}
\end{equation}
 is the lowering operating annihilating the ground state $\hat{b}\vert 0\rangle=0$ and  $[\hat{b}, \hat{b}^\dagger]=1$. This classical Hilbert space is supposed to replace 2-d plane in the commutative quantum mechanics. The corresponding quantum Hilbert space is then identified with the space of Hilbert-Schmidt operators, which are essentially trace-class bounded set of operators acting on $\mathcal{H}_{c}$. Denoting the elements of quantum Hilbert space by $\vert \psi)$ we can write 
\begin{equation}
\mathcal{H}_{q} = \{\vert\psi),Tr_{c}(\psi^{\dagger}\psi)<\infty\}.\label{i4}
\end{equation}
where subscript 'c' refers to the trace over classical Hilbert space. This norm is related to the inner product in this $\mathcal{H}_{q}$, which is defined as 
\begin{equation}
(\phi\vert\psi) = Tr_{c}(\phi^{\dagger}\psi) \label{i5}
\end{equation}
for $\vert\phi),\vert\psi) \in \mathcal{H}_q $. Using this operator formulation, the authors in \cite{b7} have obtained the exact spectrum of non-commutative harmonic oscillator. \footnote{The earlier works involving harmonic oscillators on a noncommutative plane employed star product \cite{b8} and also Seiberg-Witten map \cite{b9}\cite{b10} in presence of background electromagnetic field. (see also \cite{b11}\cite{b12} for noncommutative Landau problem)}It is however not clear how the different basis states in $\mathcal{H}_{c}$ (\ref{i2})   responds to canonical (symplectic) transformation Sp(4,R). A related issue is the nature of ground state for the non-commutative harmonic oscillator and its relation to the vacuum state $\vert 0\rangle\langle 0\vert \in \mathcal{H}_{q}$. Particularly, since the Hamiltonian of the harmonic oscillator has two independent  parameter, viz mass $\mu$ and angular frequency $\omega$ having the dimension of  length-inverse, it is expected that there will be some favoured value 
of these parameters where $\vert 0\rangle\langle 0\vert $ will  also correspond to the ground state of non-commutative harmonic oscillator. Furthermore, since a 2D isotropic oscillator Hamiltonian in the commutative plane can have a manifest SO(4) symmetry(after appropriate 
scaling of phase -space variables), it will be interesting to study the presence/absence of such a symmetry for the corresponding non-commutative harmonic oscillator. Further, since the Euclidean Lorentz group SO(4) splits in to $SU(2)\otimes SU(2)$ we can focus our attention to one of the SU(2) group itself, where each generator generates simultaneous  SO(2) rotations in a pair of two orthogonal planes. Given that the non-commutative coordinates in higher dimension, say, in three dimension fail to transform covariantly under SO(3) rotation \cite{b13}, in contrast to the commuting position like operator $\hat{X}^{c}_{i} = \frac{1}{2}(\hat{X}^{L}_{i}+\hat{X}^R_i)$ ($\hat{X}^{L/R}_i$ being the left/right actions, defined more precisely in the next section), it becomes imperative to check the same in this 2D case also under the SU(2) generators constructed $\textit{a la}$ Schwinger by using the fact that $\mathcal{H}_{q}$ can be basically identified with $\mathcal{H}_{c}\otimes\tilde{\mathcal{H}}_{c}$ ($\tilde{
\mathcal{H}}_{c}$ indicates dual space), although both $\hat{X}_{i}$ and $\hat{X}^{c}_{i}$ transform covariantly under SO(2) rotation in Moyal plane. Furthermore the spectrum of the non-commutative harmonic oscillator as was obtained in \cite{b7} is written in terms of ``bare'' parameters $\mu$ and $\omega$ occurring in the Hamiltonian and the corresponding observables parameters were not identified. In this paper we try to identify the observable parameters and try to compare with the corresponding commuting harmonic oscillator in regard to the preservation of Schwinger's SU(2) symmetry and also the time reversal symmetry.\\ 

The plan of the paper is as follows. In the next section (II) we briefly review how the representations of position and momentum operators acting on $\mathcal{H}_q$ (\ref{i4}) and satisfying non-commutative Heisenberg algebra can be obtained. In section III we construct Schwinger's SU(2) generators for both commutative and non-commutative plane in 2D and identify the commutative ``position-like" coordinates transforming covariantly under SU(2). We then consider the commutative harmonic oscillator and write its Hamiltonian in a manifestly SU(2) invariant form and obtain its spectrum in section IV. In the following section V, we consider the non-commutative oscillator and investigate the role of SU(2) symmetry in obtaining the spectrum. We then discuss the time-reversal symmetry in section VI. Finally we conclude in section VII.

\section{Representation of Non-commutative Heisenberg algebra in 2-D} \label{section-1}
Having introduced classical and quantum Hilbert spaces (\ref{i2},\ref{i4}) in the previous section, we now discuss briefly following \cite{b6,b7}, about the representation of non-commutative Heisenberg algebra.\\
If $\hat{X}_i$ and $\hat{P}_i$ are the representations of the operators $\hat{x}_i$ and the conjugate momentum respectively acting on $\mathcal{H}_q$, then a unitary representation is obtained by the following action:
\begin{equation}
 \hat{X}_i \psi = \hat{x}_i \psi, ~~~ \hat{P}_i \psi = \frac{1}{\theta}\epsilon_{ij}[\hat{x}_j, \psi]=\frac{1}{\theta}\epsilon_{ij}\left(\hat{X}_j^L-\hat{X}_j^R\right)\psi, \label{b4}
\end{equation}
where $\hat{X}_i^L$ and $\hat{X}_i^R$ refer to left and right action respectively i.e. $\hat{X}_i^L\psi=\hat{x}_i\psi$ and $\hat{X}_i^R\psi=\psi\hat{x}_i$. Note that, we have taken the action of $\hat{X}_i$ to be left action $(\hat{X}_i\equiv \hat{X}_i^L)$ by default and the momentum operator is taken to act adjointly. This ensures that $\hat{X}_i$ and $\hat{P}_i$ satisfy non-commutative Heisenberg algebra:
\begin{equation}
[\hat{X}_i^L , \hat{X}_j^L]\equiv [\hat{X}_i , \hat{X}_j]= i\theta\epsilon_{ij} ~;~ [\hat{X}_i , \hat{P}_j]= i\delta_{ij} ~;~ [\hat{P}_i , \hat{P}_j]=0
\end{equation}
It can be checked easily that the right action satisfies 
\begin{equation}
[\hat{X}_i^R , \hat{X}_j^R] = -i\theta\epsilon_{ij} ~;~ [\hat{X}_i^L , \hat{X}_j^R] =0
\end{equation}
The corresponding right actions of annihilation operator is introduced in an analogous manner:
\begin{equation}
\mathcal{H}_q\ni|m\rangle\langle n| = \frac{1}{\sqrt{n!}}|m\rangle\langle 0|\hat{B}^n \equiv (\hat{B}_R)^n \left(\frac{1}{\sqrt{n!}} |m\rangle\langle 0| \right) \label{e4}
\end{equation}
so that for any $\psi\in \mathcal{H}_q$, one defines $\hat{B}_L\psi=\hat{B}\psi$ and $\hat{B}_R\psi=\psi\hat{B}$ and similarly for $\hat{B}^\ddagger_{L/R}$, where $\hat{B}$ is the representation of $\hat{b}$ (\ref{i3}) acting on $\mathcal{H}_q$ and is given by $\hat{B} = \frac{1}{\sqrt{2\theta}}(\hat{X}_1 +i\hat{X}_2)$. Note that we are using $\dag$ and $\ddag$ to denote Hermitian conjugation over $\mathcal{H}_c$ and $\mathcal{H}_q$ respectively.
\section{Angular momentum}
We are now going to study angular momentum using Schwinger's representation. First we will discuss angular momentum operators in commutative 2D case with the help of two decoupled simple harmonic oscillators. Then we will move to non-commutative case where we will see how the roles of angular momentum operators gets interchanged with the operators in commutative case.
\subsection{Schwinger's representation of Angular momentum}
Schwinger introduced a method to express a general angular momentum in quantum mechanics in terms of the creation ($\hat{a}_\alpha^\dagger$) and annihilation ($\hat{a}_\alpha$) operators of a pair of independent harmonic oscillators satisfying:
\begin{equation}
[\hat{a}_\alpha , \hat{a}_\beta] = 0 = [\hat{a}^\dagger _\alpha , \hat{a}^\dagger _\beta] ~\textsf{and}~   [\hat{a} _\alpha , \hat{a}^\dagger _\beta] = \delta _{\alpha \beta} ~~ \forall \alpha , \beta = 1,2
\end{equation}
With the help of these operators the number operator $\hat{N}$ and angular momentum operator $\hat{\vec{J}}$ is defined as follows:
\begin{equation}
\hat{N} = \hat{a}^\dagger _\alpha \hat{a}_\alpha
\end{equation} 
\begin{equation} \label{schwinger}
\hat{\vec{J}} = \frac{1}{2} \hat{a}^\dagger _\alpha \lbrace\vec{\sigma}\rbrace_{\alpha\beta} \hat{a}_\beta ~;~ 
[\hat{J}_i , \hat{J}_j] = i\epsilon_{ijk} \hat{J}_k,
\end{equation}
with $\vec{\sigma}$ being our usual Pauli matrices.

\subsection{Role of Angular momentum operators in commutative 2-D plane}
In the commutative case let us choose our basis for a pair of decoupled harmonic oscillators which can be regarded as $|m\rangle\otimes |n\rangle\in \mathcal{H}_c\otimes\mathcal{H}_c$. We define the  creation operators $\hat{a}_1^\dagger$, $\hat{a}_2^\dagger$ as $(\hat{a}^\dagger\otimes 1)$ and $(1\otimes \hat{a}^\dagger)$ respectively, and their Hermitian conjugates as annihilation operators. Now using Schwinger's prescription (\ref{schwinger}) we obtain the three angular momentum operators as
\begin{equation}
\hat{J}_1=\frac{1}{2}\left(\hat{a}_2 ^\dagger \hat{a}_1 + \hat{a}_1 ^\dagger \hat{a}_2 \right),~~\hat{J}_2=\frac{i}{2}\left(\hat{a}_2^\dagger \hat{a}_1 - \hat{a}_1^\dagger \hat{a}_2 \right)~~\textsf{and}~~\hat{J}_3=\frac{1}{2}\left(\hat{a}_1^\dagger \hat{a}_1 -\hat{a}_2^\dagger \hat{a}_2 \right) \label{a6}
\end{equation}
such that they satisfy the $su(2)$ algebra 
\begin{equation}
\left[\hat{J}_i , \hat{J}_j \right]= i\epsilon _{ijk}\hat{J}_k 
\end{equation}
in which $\hat{J}_3$ and the Casimir $\hat{\vec{J}}^2$ operators satisfy following eigen-value equations:
\begin{eqnarray}
\hat{J}_3\big(\big|m\big>\otimes\big|n\big>)&=&j_3\big(\big|m\big>\otimes\big|n\big>\big), ~~~ \text{where} ~ j_3= \frac{1}{2}(m-n) \label{a9} \\
\hat{\vec{J}}^2\big(\big|m\big>\otimes\big|n\big>)&=&j(j+1)\big(\big|m\big>\otimes\big|n\big>),~~\text{where}~~j=\frac{1}{2}(m+n) \label{b1}
\end{eqnarray}
Let us consider the Hamiltonian of two independent harmonic oscillators with same mass $\mu$ and frequency $\omega$ as follows:
\begin{equation}
\hat{H} = \frac{1}{2}\mu\omega^2 \left(\hat{X}_1^2+\hat{X}_2^2\right)+\frac{1}{2\mu}\left(\hat{P}_1^2+\hat{P}_2^2\right)=\frac{\omega}{2}\left(\frac{\hat{\vec{P}}^2}{\mu\omega} + \mu\omega\hat{\vec{X}}^2\right) \label{a1}
\end{equation}
This can be identified with the 2D harmonic oscillator on a commutative plane. We can then write the 4D phase space variables in terms of the respective ladder operators as:
\begin{eqnarray}
\hat{X}_{\alpha}&=&\frac{1}{\sqrt{2\mu\omega}}\left(\hat{a}_{\alpha}+\hat{a}_{\alpha}^\dagger\right) \label{a2}\\ 
\hat{P}_{\alpha}&=&i\sqrt{\frac{\mu\omega}{2}}\left(\hat{a}_{\alpha}^\dagger -\hat{a}_{\alpha}\right)~~\forall~~\alpha= 1,2 \label{a3}
\end{eqnarray}
Now we perform following canonical transformation$\big($ which would preserve the commutation relation $[\hat{X}_{\alpha},\hat{P}_{\beta}]=i\delta_{\alpha\beta}\big)$:
$$\hat{P}_{\alpha}\rightarrow\hat{p}_{\alpha}=\frac{\hat{P}_{\alpha}}{\sqrt{\mu\omega}}~~\textsf{and}~~\hat{X}_{\alpha}\rightarrow\hat{x}_{\alpha}=\sqrt{\mu\omega}\hat{X}_{\alpha}$$
With this our Hamiltonian becomes $\hat{H}=\frac{\omega}{2}\left(\hat{x}_1^2+\hat{x}_2^2+\hat{p}_1^2+\hat{p}_2^2\right)$, which clearly enjoys SO(4) symmetry in the 4D phase space.\\
To understand the action of the Schwinger's angular momentum operators (\ref{a6}) let us calculate the following commutation relations:
\begin{equation}
[\hat{J}_3,\hat{x}_{\alpha}]=\frac{i}{2}\left(\delta_{\alpha 2}\hat{p}_2-\delta_{\alpha 1}\hat{p}_1\right)~~\textsf{and}~~ [\hat{J}_3,\hat{p}_{\alpha}]=\frac{i}{2}\left(\delta_{\alpha 1}\hat{x}_1-\delta_{\alpha 2}\hat{x}_2\right) \label{f1}
\end{equation}
\begin{equation}
[\hat{J}_1,\hat{x}_{\alpha}]=\frac{-i}{2}\left(\delta_{\alpha 1}\hat{p}_2+\delta_{\alpha 2}\hat{p}_1\right)~~\textsf{and}~~ [\hat{J}_1,\hat{p}_{\alpha}]=\frac{i}{2}\left(\delta_{\alpha 1}\hat{x}_2+\delta_{\alpha 2}\hat{x}_1\right)
\end{equation}
\begin{equation}
[\hat{J}_2,\hat{x}_{\alpha}]=\frac{i}{2}\epsilon_{\alpha\beta}\hat{x}_{\beta}~~\textsf{and}~~[\hat{J}_2,\hat{p}_{\alpha}]=\frac{i}{2}\epsilon_{\alpha\beta}\hat{p}_{\alpha}; ~~\textsf{with}~~ \alpha ,\beta =1,2. \label{f2}
\end{equation}
From the above relations we can conclude that $\hat{J}_3$ generates simultaneous SO(2) rotation in $x_1 p_1$ and $x_2 p_2$ planes. Like-wise in case of $\hat{J}_1$ the rotation occurs in $x_1 p_2$ and $x_2 p_1$ planes, where as for $\hat{J}_2$ the rotation occurs in $x_1 x_2$ and $p_1 p_2$ planes. \\
This is however a part of the $SO(4)$ symmetry only, as the $SU(2)$ symmetry generated by this $\hat{\vec{J}}$'s corresponds to one of the $su(2)$'s in the decomposition of  so(4) Lie algebra as $su(2) \oplus su(2)$. To get the other $su(2)$, one has to just flip the sign of one of the momenta components, say of $p_2$. 
\subsection{Schwinger's Angular momentum operators in non-commutative 2-D plane}
As we have discussed in section-1, the quantum Hilbert space(\ref{i4}) comprises of Hilbert-Schmidt operators, and therefore any generic Hilbert-Schmidt operator can be written as: 
$$ \big|\Psi\big)=\sum_{m,n}C_{mn}|m\rangle\langle n|\in\mathcal{H}_q $$
The $\mathcal{H}_q$ can be identified with $\mathcal{H}_c \otimes \tilde{\mathcal{H}}_c$ , where $\tilde{\mathcal{H}}_c$ is the dual of $\mathcal{H}_c$. Since, there is a one-to-one map between the basis $|m\rangle\otimes|n\rangle$ and $|m\rangle\otimes\langle n|$, the Hilbert spaces, $\text{span} \lbrace|m\rangle\otimes|n\rangle\rbrace$=$\mathcal{H}_c\otimes\mathcal{H}_c$ and $\mathcal{H}_q$(\ref{i4}) are isomorphic. In order to obtain the angular momentum operators acting on $\mathcal{H}_q$ i.e the counterpart of the expressions in (\ref{a6}), let us replace $\hat{a}_1$ with $\hat{B}_L$ and $\hat{a}^\dagger _2$(and not $\hat{a}_2$) with $\hat{B}_R$ and their respective Hermitian conjugates. Here the operators $\hat{B}_{L/R}$ and $\hat{B}_{L/R}^\ddagger$ are the operators on quantum Hilbert space $\mathcal{H}_q$ such that $\hat{B}_L=\hat{b}\otimes 1$ and $\hat{B}_R=1\otimes \hat{b}_R$(\ref{e4}). Hence we get:
\begin{equation}
\hat{J}_1=\frac{1}{2}\left( \hat{B}_R \hat{B}_L + {\hat{B}_L}^\ddagger {\hat{B}_R}^\ddagger \right),~~\hat{J}_2=\frac{i}{2}\left( \hat{B}_R \hat{B}_L - {\hat{B}_L}^\ddagger {\hat{B}_R}^\ddagger \right)~~\textsf{and}~~ \hat{J}_3=\frac{1}{2}\left({\hat{B}_L}^\ddagger \hat{B}_L -  \hat{B}_R {\hat{B}_R}^\ddagger \right)
\end{equation}
Clearly they satisfy the su(2) algebra 
\begin{equation}
\left[ \hat{J}_i , \hat{J}_j \right]= i\epsilon _{ijk} \hat{J}_k,
\end{equation}
in which $\hat{J}_3$ satisfies the eigen-value equation:
\begin{equation}
\hat{J}_3 |m\rangle\langle n| = j_3 |m\rangle\langle n|, ~~~ \text{where} ~ j_3= \frac{1}{2}(m-n) \label{b2}
\end{equation}
and the corresponding Casimir operator
\begin{equation}
\hat{J}^2=\frac{1}{4}\left({\hat{B}_L}^\ddagger \hat{B}_L +  \hat{B}_R {\hat{B}_R}^\ddagger \right)\left({\hat{B}_L}^\ddagger \hat{B}_L +  \hat{B}_R {\hat{B}_R}^\ddagger + 2 \right), \label{b3}
\end{equation}
satisfying following eigen-value equation:
\begin{equation}
\hat{J}^2 |m\rangle\langle n| = j(j+1) |m\rangle\langle n|, ~~~ \text{where} ~ j= \frac{1}{2}(m+n) \label{d3}
\end{equation}
These pair of eigen-values $j_3(\ref{b2})~and~j(j+1)(\ref{b3})$ are exactly identical to the ones in (\ref{a9}) and (\ref{b1}). This suggests that both the basis states $\big|m\big>\otimes\big|n\big>\in\mathcal{H}_c\otimes\mathcal{H}_c$ and $\big|m,n\big)\equiv\big|m\big>\otimes\big<n\big|\in\mathcal{H}_c\otimes \tilde{\mathcal{H}}_c$ can be labeled alternatively as $\big|j,j_3\big)=\big|m,n\big)$ where $j=\frac{1}{2}(m+n)$ and $j_3=\frac{1}{2}(m-n)$. This is illustrated in following diagram(figure 1). We see from the diagram that all the discrete points with integer-valued coordinates $(m,n)$ with $m,n\geq0$ in the first quadrant represent a basis element $|m,n)\equiv|m\rangle\langle n|$ of the Hilbert space of states $\mathcal{H}_q$. Hence we find that the Hilbert space gets split up into states of constant $j$ value lying on the straight line running parallel to $j_3$ axis e.g. $j=\frac{1}{2}$ line, $j=1$ line and so on, and $j_3$ takes values within the interval $-j\leq j_3\leq j$ \\
We can then construct the usual ladder operators $\hat{J}_{\pm}=\hat{J}_1\pm i\hat{J}_2$, connecting all states belonging to fixed $'j'$ subspace of $(2j+1)$ dimension and the extremal points are annihilated by $\hat{J}_{\pm}$. This is just what happens in usual quantum mechanics.\\ 
We can now express the position and momentum operators in terms of $\hat{B}_L , {\hat{B}_L}^\ddagger,$ $\hat{B}_R$ and ${\hat{B}_R}^\ddagger $. By making use of equation (\ref{b4}), the definition  $\hat{B}_{L/R}=\frac{1}{\sqrt{2\theta}}(\hat{X}_1^{L/R}+i\hat{X}_2^{L/R})$ \cite{b5} and its Hermitian conjugate, and the adjoint action of momenta(\ref{b4}) we get:  
\begin{equation}
\hat{X}_1^L = \sqrt{\frac{\theta}{2}}(\hat{B}_L+ \hat{B}_L ^{\ddagger})
\end{equation}
\begin{equation}
\hat{X}_2^L = i\sqrt{\frac{\theta}{2}}(\hat{B}_L ^{\ddagger} - \hat{B}_L)
\end{equation}
\begin{equation}
\hat{P}_1 = \frac{i}{\sqrt{2\theta}} \left({\hat{B}_L}^\ddagger - \hat{B}_L   - {\hat{B}_R}^\ddagger + \hat{B}_R \right) \label{b9}
\end{equation}
\begin{equation}
\hat{P}_2 = \frac{1}{\sqrt{2\theta}} \left({\hat{B}_R}^\ddagger + \hat{B}_R -{\hat{B}_L}^\ddagger - \hat{B}_L \right)
\end{equation}
Further, the commuting coordinates \cite{b5} introduced as
\begin{equation}
\hat{X}_i^c=\frac{1}{2}\left(\hat{X}_i^L+\hat{X}_i^R\right)= \hat{X}_i + \frac{\theta}{2}\epsilon _{ij} \hat{P}_j  \label{f3}
\end{equation}
satisfying $[\hat{X}_i^c,\hat{X}_j^c]=0$, can be expressed like-wise as: 
\begin{equation}
\hat{X}_1 ^c = \frac{1}{2}\sqrt{\dfrac{\theta}{2}} \left( \hat{B}_R + \hat{B}_L + {\hat{B}_L}^\ddagger + {\hat{B}_R}^\ddagger \right)
\end{equation}
\begin{equation}
\hat{X}_2 ^c = \frac{i}{2}\sqrt{\dfrac{\theta}{2}} \left({\hat{B}_L}^\ddagger - \hat{B}_L   + {\hat{B}_R}^\ddagger -\hat{B}_R \right) \label{c1}
\end{equation}
To see the similarity of the action of Schwinger's angular momentum operators with their commutative counterpart(\ref{f1}-\ref{f2}), we first perform following canonical transformation, to construct dimensionless phase space variables: 
\begin{figure}
\includegraphics[width=12cm,height=8cm]{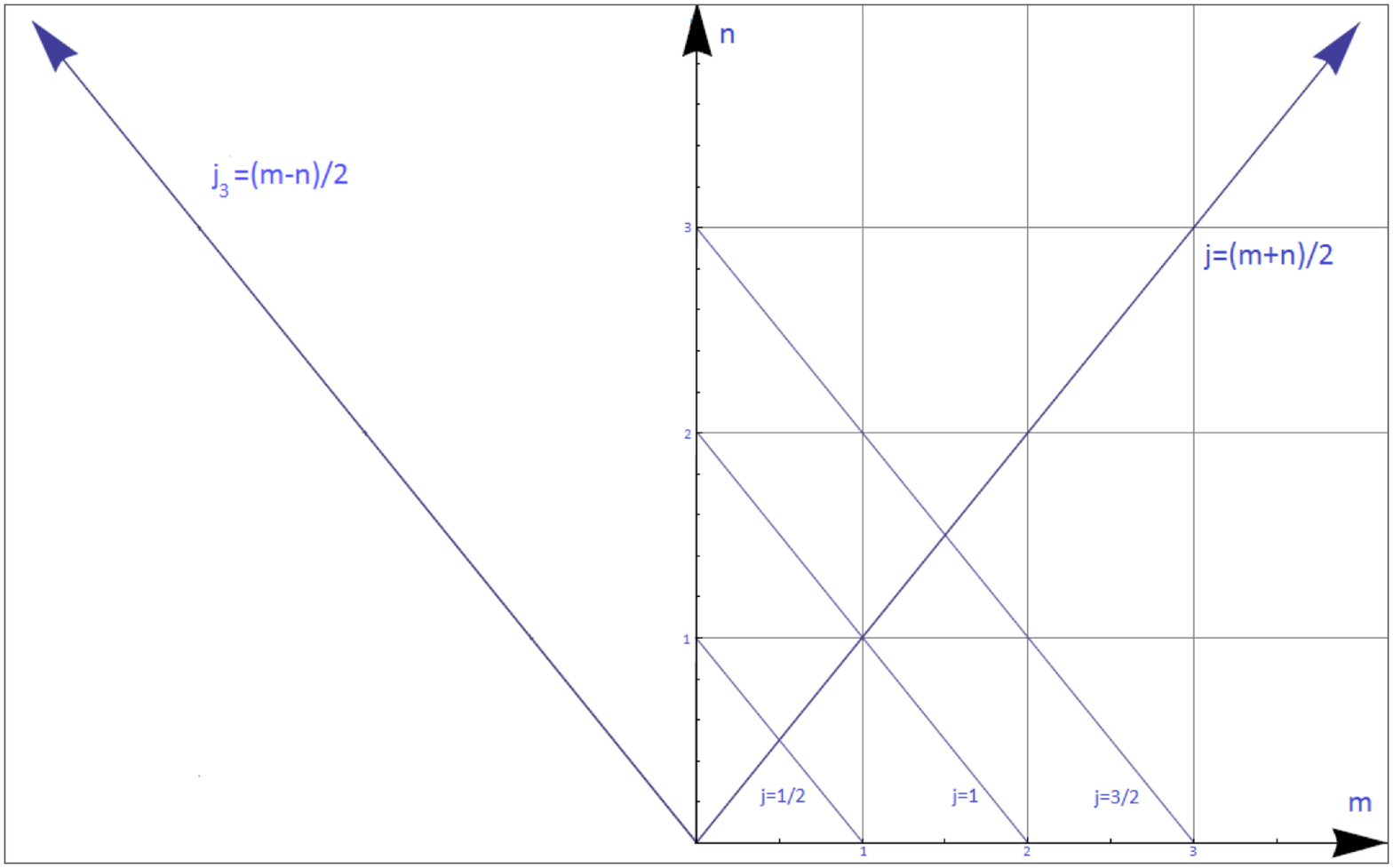}
\end{figure}
\begin{equation}
\hat{X}_i^c\rightarrow\hat{x}_i^c= \frac{\hat{X}_i^c}{\sqrt{\theta}} ~~\textsf{and}~~ \hat{P}_i\rightarrow\hat{p}_i= \sqrt{\theta}\hat{P}_i ~~~~\forall ~i=1,2  \label{b8}
\end{equation}
Then we calculate following commutation relations:
\begin{equation}
[\hat{x}_i^c,\hat{J}_1] = \frac{i}{2}\left(\delta_{i1}\hat{p}^{\prime}_1 - \delta_{i2}\hat{p}^{\prime}_2\right) ~ ~\textsf{and} ~~ [\hat{p^{\prime}_i},\hat{J}_1] = \frac{i}{2}\left(\delta_{i2}\hat{x}_2 - \delta_{i1}\hat{x}_1\right) \label{b5}
\end{equation}
\begin{equation}
[\hat{x}_i^c,\hat{J}_2] = -\frac{i}{2}\left(\delta_{i1}\hat{p}^{\prime}_2+\delta_{i2}\hat{p}^{\prime}_1\right)~ ~\textsf{and} ~~
[\hat{p}^{\prime}_i,\hat{J}_2]= \frac{i}{2}\left(\delta_{i1}\hat{x}_2 +\delta_{i2}\hat{x}_1\right)
\end{equation}
\begin{equation}
[\hat{x}_i^c,\hat{J}_3]=\frac{i}{2}\epsilon_{ij}\hat{x}_j ~ ~\textsf{and} ~~ [\hat{p}^{\prime}_i,\hat{J}_3]=\frac{i}{2}\epsilon_{ij}\hat{p}^{\prime}_j \label{b6}
\end{equation}
Here we have introduced $\hat{p}_i^{\prime}=\frac{\hat{p}_i}{2}$. So we see that the angular momentum operators $\hat{J}_i$ are responsible for inducing simultaneous SO(2) rotations in two orthogonal planes of our 4-D phase space. In fact for $\hat{J}_1$ the rotation occurs in $X_1 ^c P^{\prime}_1$ and $X_2 ^c P^{\prime}_2$ planes. In case of $\hat{J}_2$ the rotation occurs in $X_1 ^c P^{\prime}_2$ and $X_2 ^c P^{\prime}_1$ planes, where as for $\hat{J}_3$ the rotation occurs in $X_1 ^c X_2 ^c$ and $P^{\prime}_1 P^{\prime}_2$ planes. Here we can also appreciate that  roles of commutative and non-commutative angular momentum operators are exchanged in following manner (where superscript C refers to commutative case(\ref{f1}-\ref{f2}) and NC for non-commutative(\ref{b5}-\ref{b6}) case):
\begin{equation}
\hat{J}_1^{NC}\leftrightarrow\hat{J}_3^C,~~\hat{J}_2^{NC}\leftrightarrow\hat{J}_1^C~~\textsf{and}~~\hat{J}_3^{NC}\leftrightarrow\hat{J}_2^C 
\end{equation} 
Again like the commutative case this is also a part of $SO(4)$ rotation in $X_1 ^c, X_2 ^c, P_1, P_2$ space, as $\hat{J}^{NC}$'s corresponds to only one of the $su(2)$'s of $so(4) = su(2) \oplus su(2)$. In either case, this is just reminiscent of the splitting of the SL(2,C) algebra by taking suitable combinations of spatial rotations with Lorentz boost in $3+1$ dimensional Minkowski space-time, and SL(2,C) happens to be just the double cover of the Lorentz group SO(3,1),whose Euclidean version is the spin group Spin(4), which is the double cover of SO(4). Now we can obtain the rotation matrix for finite rotation $R_i(\lambda _i)$ generated by $\hat{J}_i$ in the $4D$ phase space by using above commutation relations (\ref{b5}-\ref{b6})  which is implemented by the following unitary transformation
\begin{eqnarray}
\Xi \longrightarrow \Xi^\prime&=&e^{-i\vec{\lambda}.\vec{J}} \Xi e^{i\vec{\lambda}.\vec{J}}=R(\vec{\lambda})\Xi \nonumber \\  \label{b7}
\end{eqnarray}
where the 4-component column matrix $\Xi$ comprises of phase space variables and is given by:
\begin{equation}
\Xi =\left(\hat{x_1}^c,\hat{x_2}^c,\frac{\hat{p}_1}{2},\frac{\hat{p}_2}{2}\right)^T \label{c8}
\end{equation} 
The matrix representations of the corresponding generators in the 4-dimensional representation are given by $\hat{J}_i = -i\frac{\partial R_i(\vec{\lambda})}{\partial \lambda_i}\Big|_{\vec{\lambda}=0}$. This yields
\begin{eqnarray}
\hat{J}_1 = \frac{i}{2}\begin{pmatrix}
0 & 0 & 1 & 0 \\
0 & 0 & 0 & -1 \\
-1 & 0 & 0 & 0 \\
0 & 1 & 0 & 0 \\
\end{pmatrix}
; \hat{J}_2 = \frac{i}{2}\begin{pmatrix}
0 & 0 & 0 & -1 \\
0 & 0 & -1 & 0 \\
0 & 1 & 0 & 0 \\
1 & 0 & 0 & 0 \\
\end{pmatrix}
; \hat{J}_3 = \frac{i}{2}\begin{pmatrix}
0 & 1 & 0 & 0 \\
-1 & 0 & 0 & 0 \\
0 & 0 & 0 & 1 \\
0 & 0 & -1 & 0 \\
\end{pmatrix}
\end{eqnarray}
The corresponding Casimir operator is then given by:
\begin{equation}
\hat{\vec{J}}^2 = \hat{J}_1^2+\hat{J}_2^2+\hat{J}_3^2 = \frac{1}{2}\left(\frac{1}{2}+1\right) I
\end{equation}
indicating that this corresponds to spin $\frac{1}{2}$ representation of $SU(2)\subset SO(4)$, which looks odd from a naive consideration, where it appears to be $j=\frac{3}{2}$ representation as the dimensionality is 2j+1=4. This is however easily explained by fact that the $\hat{J}'s$ are making simultaneous rotation in two different orthogonal planes through $j=\frac{1}{2}$ representation.
However, the various basis states $|j,j_3)$ in fig.1 will transform according to the $j-$th representation of Schwinger's SU(2) generators.\\
Finally, we would like to point out the important role played by the commuting but unphysical position-like operators $\hat{X}_i^c$(\ref{f3}). It is only $\hat{X}_i^c$, rather than the physical non-commuting position operators $\hat{X}_i$, that transform covariantly under SU(2). It is only under SO(2) rotation, generated by $\hat{J}_3$, that both $\hat{X}_i^c$ and $\hat{X}_i$ transform covariantly. As has been shown in \cite{b13}, that in 3D Moyal space also, it is only $\hat{X}_i^c$ that transform covariantly under SO(3) rotation, although $\hat{X}^c_i$ does not correspond to any physical observable \cite{b5}. Further, as we shall see later that this is also true for Bogoliubov transformation, where it undergoes simple scaling transformation and plays a very important role in our analysis, despite being an unphysical observable \cite{b5}.
\section{Simple Harmonic Oscillator in commutative plane} 
In commutative ($\theta=0$) case the Hamiltonian is given by(\ref{a1}):
\begin{equation}
\hat{H}_I = \frac{1}{2}\mu\omega^2\hat{\vec{X}}^2+\frac{1}{2\mu}\hat{\vec{P}}^2 \label{c6}
\end{equation}
Using the form of position and momentum operators in terms of ladder operators (\ref{a2}-\ref{a3}), we get following result:
\begin{eqnarray}
\hat{\vec{X}}^2&=&\frac{1}{2\mu\omega}\left(\hat{a}_1 \hat{a}_1+\hat{a}_1^\dagger \hat{a}_1^\dagger+\hat{a}_2 \hat{a}_2+\hat{a}_2^\dagger \hat{a}_2^\dagger\right)+\frac{1}{\mu\omega}\left(\hat{a}_1^\dagger \hat{a}_1+\hat{a}_2^\dagger \hat{a}_2+1\right) \label{a4}\\
\hat{\vec{P}}^2&=&-\frac{\mu\omega}{2}\left(\hat{a}_1 \hat{a}_1+\hat{a}_1^\dagger \hat{a}_1^\dagger +\hat{a}_2\hat{a}_2+\hat{a}_2^\dagger \hat{a}_2^\dagger\right)+\mu\omega\left(\hat{a}_1^\dagger \hat{a}_1+\hat{a}_2^\dagger \hat{a}_2+1\right), \label{a5}
\end{eqnarray}
so that our Hamiltonian (\ref{c6}) becomes:
\begin{equation}
\hat{H}_I=\omega\left(a_1^\dagger a_1+a_2^\dagger a_2+1\right),
\end{equation}
which is clearly SU(2) invariant as it commutes with angular momentum operators $\hat{J}_i$(\ref{a6}). Using our most general state $|m\rangle\otimes|n\rangle$, we get following spectrum:
\begin{equation}
\hat{H}_I|m\rangle\otimes|n\rangle=\omega(2j+1)|m\rangle\otimes|n\rangle,~~\textsf{where}~~j=\frac{1}{2}(m+n), \label{d1}
\end{equation}
which was expected as our state $|m\rangle\otimes|n\rangle$ corresponds to pair of simple harmonic oscillator (one ket for each SHO), and the energy of any SHO, say $|m\rangle$ is $\omega\left(m+\frac{1}{2}\right)$. SU(2) invariance of the Hamiltonian is also evident from above as spectrum  is given by $E(j,j_3)= \omega(2j+1) $ and is independent of $j_3$ and (2j+1) being the dimensionality of j-th subspace.
\section{Simple Harmonic oscillator in non-commutative plane}
Here we shall first analyze the unphysical Hamiltonian of simple harmonic oscillator involving the commuting coordinates $\hat{X}_i^c$ \cite{b5}, which is shown to yield the same spectrum as in the commutative case(\ref{d1}). This will then pave the way to analyze the physical oscillator involving non-commuting operators. 
\subsection{Unphysical SHO involving ($\hat{X}_i^c$)}
Let us first construct following quadratic form:
\begin{equation}
\hat{H}_1=\left(\frac{1}{\theta}(\hat{\vec{X}}^c)^2+\frac{\theta}{4}\hat{\vec{P}}^2\right), \label{e7}
\end{equation}
which can be regarded formally as a Hamiltonian of a harmonic oscillator with particular choices of mass and angular frequency parameters. But since it involves $\hat{X_i}^c$ (\ref{f3}), which is just a mathematically constructed commuting observable and is devoid of any physical interpretation \cite{b5}, this oscillator is really unphysical in nature. Now using canonical transformation (\ref{b8}) above Hamiltonian becomes:
\begin{equation}
\hat{H}_1=\left((\hat{x}_1^c)^2 +(\hat{x}_2^c)^2 +\Big(\frac{\hat{p_1}}{2}\Big)^2 +\Big(\frac{\hat{p_2}}{2}\Big)^2\right)=\Xi^T \Xi ~~\textsf{using}~~(\ref{c8}), \label{f4}
\end{equation}
which is manifestly invariant under one of the SU(2) symmetry group in the decomposition of SO(4) i.e $SO(4)=SU(2)\otimes SU(2)$ analogous to commuting case.\footnote{Actually, it is invariant under the entire SO(4) group itself; the other SU(2) symmetry is realized by flipping the sign of one of the momenta component say of $\hat{p}_2$, as was observed earlier. In our present analysis, we shall not make use of this second SU(2). We shall therefore continue to refer to the first SU(2) symmetry only.} This is evident from commutation relations (\ref{b5}-\ref{b6}) and unitary transformation (\ref{b7}).\\ 
Then (using \ref{b9}-\ref{c1}), one gets
\begin{equation}
(\hat{\vec{X}}^c)^2=(\hat{X}_1 ^c)^2+(\hat{X}_2 ^c)^2=\frac{\theta}{2}\left( {\hat{B}_L}^\ddagger \hat{B}_L +  \hat{B}_R {\hat{B}_R}^\ddagger +1+ \hat{B}_L {\hat{B}_R}^\ddagger + {\hat{B}_L}^\ddagger \hat{B}_R \right) \label{c4}
\end{equation}
\begin{equation}
\hat{\vec{P}}^2=(\hat{P}_1)^2+(\hat{P}_2)^2=\frac{2}{\theta}\left( {\hat{B}_L}^\ddagger \hat{B}_L +  \hat{B}_R {\hat{B}_R}^\ddagger +1- \hat{B}_L {\hat{B}_R}^\ddagger - {\hat{B}_L}^\ddagger \hat{B}_R \right), \label{c5}
\end{equation}
so that our Hamiltonian (\ref{f4}) becomes:
\begin{equation}
\hat{H}_1 = \frac{1}{2}\left(\frac{2}{\theta}(\hat{\vec{X}}^c)^2 + \frac{\theta}{2}\hat{\vec{P}}^2\right)= \left( {\hat{B}_L}^\ddagger \hat{B}_L +  \hat{B}_R {\hat{B}_R}^\ddagger +1\right) \label{c2}
\end{equation}
Hence for a general basis state $|m\rangle\langle n|$ we get following spectrum:
\begin{equation}
\hat{H}_1 |j,j_3)= \hat{H}_1|m \rangle\langle n |=(m+n+1)|m \rangle\langle n |=(2j+1)|j,j_3), ~~\textsf{using}~~(\ref{d3}) \label{2a}
\end{equation}
and the spectrum can be read-off easily, yielding just $(2j+1)$.\\
 We then consider following Hamiltonian, which is a slight variant for the non-physical planar harmonic oscillator(\ref{e7}), by introducing mass $(\mu)$ and angular frequency $(\omega)$: 
\begin{equation}
\hat{H}_2 = \frac{1}{2\mu}\hat{\vec{P}}^2 + \frac{1}{2} \mu\omega ^2 (\hat{\vec{X}}^c)^2 \label{c3} 
\end{equation}
Naively we expect that this Hamiltonian to be equivalent to $\hat{H}_I$(\ref{c6}), however this is not the case because there is a major difference in the construction of actually commuting (\ref{c6}) and mathematically constructed commuting $\hat{X}_i^c$ coordinates. The annihilation operators $a_1$ and $a_2$ occurring in(\ref{a2}) act on the left and right slots separately, annihilating the vacuum $|0\rangle\otimes|0\rangle$. In this case we can define creation and annihilation operators in an analogous manner to(\ref{a2} and \ref{a3}) as:
\begin{equation}
\hat{C}_i^{\dagger}=\frac{1}{\sqrt{2\mu\omega}}\left(\mu\omega\hat{X}_i^c-i\hat{P}_i\right),~~\hat{C}_i=\frac{1}{\sqrt{2\mu\omega}}\left(\mu\omega\hat{X}_i^c+i\hat{P}_i\right) ~~\forall~~ i=1,2, \label{e5}
\end{equation}
which, however involves $\hat{X}_i^c$ and $\hat{P}_i$ acting simultaneously in the left and right slots. One can see that the relevant ground state here i.e. $|0\rangle\otimes\langle 0|$ is annihilated by $C_i$ only under a special choice of parameters $\mu$ and $\omega$, which we call as a $\textit{critical point}:$
\begin{equation}
\mu_0=\frac{\omega_0}{2}=\frac{1}{\sqrt{\theta}} \label{c7}
\end{equation}
Also we notice that at this critical point $\hat{H}_2$(\ref{c3}) reduces to $\hat{H}_1$(\ref{e7}), up to an overall constant: $\hat{H}_2 = \frac{1}{\sqrt{\theta}}\hat{H}_1$ and furthermore, the ladder operators $\hat{C}_i$ and $\hat{C}_i^\dagger$ (\ref{e5}) reduce to the following linear combination of operators $\hat{B}_L$ and $\hat{B}_R^\ddagger$:

\begin{equation}
\hat{C}_1=\frac{1}{\sqrt{2}}\left(\hat{B}_L+\hat{B}_R^\ddagger\right);~~\hat{C}_2=\frac{-i}{\sqrt{2}}\left(\hat{B}_L-\hat{B}_R^\ddagger\right) \label{f6}
\end{equation} 

It is therefore interesting to see how our Hamiltonian(\ref{c3}) behaves under above condition(\ref{c7}). We find that:
\begin{equation}
\hat{H}_2=\omega_0\left(\frac{1}{\theta}(\hat{\vec{X}}^c)^2+\frac{\theta}{4}\hat{\vec{P}}^2\right)
=\omega_0\left({\hat{B}_L}^\ddagger \hat{B}_L +  \hat{B}_R {\hat{B}_R}^\ddagger +1\right), \label{c9}
\end{equation}
so that its action on $|m,n)\equiv |m\rangle\langle n| \equiv |j,j_3)$ gives
\begin{equation}
\hat{H}_2|m\rangle\langle n|=\omega_0(2j+1)|m\rangle\langle n| \label{f5}
\end{equation}
yielding the spectrum as 
\begin{equation}
E(j,j_3)= \omega_0 (2j+1) \label{N}
\end{equation} 
matching with the spectrum of commutative case(\ref{d1}), as was expected. It should be mentioned here that the contrasting situation in the non-commutative context from the commutative case discussed in the previous section becomes clear here. The mass and frequency parameter $\mu$ and $\omega$ requires to be compatible with each other, as in (\ref{c7}) and also with the natural mass scale introduced by $(\frac{1}{\sqrt{\theta}})$. \\
Now in order to calculate the spectrum of the Hamiltonian $\hat{H}_2$(\ref{c3}) for arbitrary value of parameters $\mu$ and $\omega$, we write it in following form by using(\ref{c4}-\ref{c5}):
\begin{equation}
\hat{H}_2=\alpha \left( {\hat{B}_L}^\ddagger \hat{B}_L + {\hat{B}_R}^\ddagger \hat{B}_R \right) +\beta \left( {\hat{B}_L}^\ddagger \hat{B}_R +  {\hat{B}_R}^\ddagger  \hat{B}_L \right) \label{eq60}
\end{equation}
where 
\begin{equation}
\alpha = \frac{\mu\omega^2 \theta}{4} + \frac{1}{\mu\theta} ~~ \text{and}~~ \beta =\frac{\mu\omega^2 \theta}{4} -\frac{1}{\mu\theta} \label{C}
\end{equation}
Clearly Hamiltonian contains off-diagonal terms. To diagonalize the Hamiltonian, let us introduce the new set of operators $\hat{B}_L^\prime$ and $\hat{B}_R^\prime$, which are related to un-primed ones via Bogoliubov transformation as:
\begin{equation}
\begin{pmatrix}
\hat{B}_L^\prime\\\hat{B}_R^\prime
\end{pmatrix}=\begin{pmatrix}
\cosh\phi & \sinh\phi \\
\sinh\phi & \cosh\phi 
\end{pmatrix}\begin{pmatrix}
\hat{B}_L\\\hat{B}_R
\end{pmatrix} \label{d4}
\end{equation}
This ensures $[\hat{B}_L^{\prime},\hat{B}_L^{\prime \ddagger}]=-[\hat{B}_R^{\prime},\hat{B}_R^{\prime \ddagger}]=1$ like their un-primed counterparts. With above transformation our Hamiltonian becomes:
\begin{eqnarray} \nonumber
\hat{H}_2 =&&\left[ \alpha  \left( \cosh ^2 \phi +\sinh ^2 \phi \right)- 2\beta \sinh \phi \cosh \phi  \right] \left(\hat{B}_L^{\prime\ddagger} \hat{B}_L^\prime + \hat{B}_R^\prime \hat{B}_R^{\prime\ddagger} \right) \\ \nonumber
&+& \left[ \beta \left( \cosh ^2 \phi +\sinh ^2 \phi \right)- 2\alpha \sinh \phi \cosh \phi  \right] \left( \hat{B}_L^{\prime\ddagger} \hat{B}_R^\prime + \hat{B}_R^{\prime\ddagger} \hat{B}_L^\prime \right) \\ \nonumber
&+& 2\alpha \sinh ^2 \phi - 2\beta \sinh \phi \cosh \phi + \alpha 
\end{eqnarray} 
For diagonalizing the Hamiltonian, let us set the coefficient of the non diagonal term zero i.e.
\begin{equation}
[\beta \left( \cosh ^2 \phi +\sinh ^2 \phi \right)- 2\alpha \sinh \phi \cosh \phi ]=0,
\end{equation}
which gives
\begin{equation}
\coth \phi + \tanh \phi =  \frac{2\alpha}{\beta} \label{d6}
\end{equation}
and the Hamiltonian becomes:
\begin{equation}
\hat{H}_2= 2\beta \sinh \phi \cosh \phi \left[ \left( \frac{\alpha}{\beta} \right)^2 -1 \right]\left( \hat{B}_L^{\prime\ddagger} \hat{B}_L^\prime + \hat{B}_R^\prime \hat{B}_R^{\prime\ddagger} +1 \right) \label{e1}
\end{equation}
We can now determine the spectrum easily by acting above Hamiltonian on an appropriate basis $|m\rangle^\prime\langle n|^\prime$ which is related to older basis $|m\rangle\langle n|$ by a unitary transformation, as we will show later.\\
Now we proceed to show that the Bogoliubov transformation used above is equivalent to a certain canonical transformation. To achieve this we expand the ladder operators in matrix equation(\ref{d4}) in terms of position operators acting from left and right to get the following set of equations:
\begin{eqnarray}
\hat{X}_1^{\prime L} &=&\cosh\phi\hat{X}_1^L+\sinh\phi\hat{X}_1^R~;~\hat{X}_2^{\prime L}=\cosh\phi\hat{X}_2^L+\sinh\phi\hat{X}_2^R \nonumber\\
\hat{X}_1^{\prime R}&=&\cosh\phi\hat{X}_1^R+\sinh\phi\hat{X}_1^L~;~\hat{X}_2^{\prime R}=\cosh\phi\hat{X}_2^R+\sinh\phi\hat{X}_2^L, \label{d5}
\end{eqnarray}
which when re-expressed in terms of $\hat{X}_i^c$(\ref{f3}) and $\hat{P}_i$(\ref{b4}), takes the following simple form:
\begin{equation}
\hat{X}_i^{\prime c}=e^\phi\hat{X}_i^c~;~~\hat{P}_i^\prime=e^{-\phi}\hat{P}_i \label{d7}
\end{equation}
Clearly, this is a canonical transformation in the $(\hat{X}_i^c,\hat{P}_j)$ space, where $\hat{X}_i^c$'s are scaled and $\hat{P}_i$'s are de-scaled. The scaling factor $e^\phi$ can be determined easily by solving the quadratic equation in $\tanh\phi$(\ref{d6}). The two roots are evidently reciprocal to each other and choosing one smaller than unity yields the value  
\begin{equation}
\tanh\phi=\frac{1}{\beta}(\alpha-\omega) \label{B}
\end{equation} 
Putting back the value of $\alpha$ and $\beta$ we finally get:
\begin{equation}
e^{\phi}=\sqrt{\frac{\mu\omega\theta}{2}}, \label{d9}
\end{equation}
which clearly reduces to unity at the critical point(\ref{c7}). This has some points of contact with the Eq.(18) in \cite{b8}. It can be easily checked that under this scaling transformation, the spatial non-commutative parameter preserves its value: $[\hat{X}_i,\hat{X}_j]= [\hat{X}_i^{\prime},\hat{X}_j^{\prime}]=i\theta\epsilon_{ij}$. It is interesting to note here that the canonical relation(\ref{d7}) is equivalent to following unitary transformation:
\begin{equation}
{\hat{X^\prime}_i}^c=e^\phi\hat{X}_i^c=e^{-\frac{i}{2}\phi\hat{D}}\hat{X}_i^c e^{\frac{i}{2}\phi\hat{D}}, \label{d8}
\end{equation}
where $\hat{D}=\frac{1}{2}(\hat{X}^c_i\hat{P}_i+\hat{P}_i\hat{X}_i^c)=i(\hat{B}_L^\ddagger \hat{B}_R-\hat{B}_L \hat{B}_R^\ddagger)$ is the dilatation operator. Correspondingly, one finds
\begin{equation}
\hat{B}^{\prime}_L\pm\hat{B}^{\prime}_R=e^{-\frac{i}{2}\phi\hat{D}}(\hat{B}_L\pm \hat{B}_R) e^{\frac{i}{2}\phi\hat{D}} = e^{\pm\phi}(\hat{B}_L\pm \hat{B}_R)\label{z1}
\end{equation} 
and the basis states in $\mathcal{H}_q$ undergo following transformation:
\begin{equation}
|m\rangle\langle n|\rightarrow |m\rangle^\prime\langle n|^\prime=e^{i\phi\hat{D}}|m\rangle\langle n|=e^{\phi(\hat{B}_R^\ddagger \hat{B}_L-\hat{B}_L^\ddagger \hat{B}_R)}|m\rangle\langle n| \label{e6}
\end{equation}
Hence instead of doing Bogoliubov transformation, we could have equivalently done the canonical transformation(\ref{d7}) to re-write the Hamiltonian(\ref{c3}) in a manifestly SU(2) invariant form in the primed frame as:
\begin{equation}
\hat{H}_2=\omega\left(\frac{1}{\theta}(\hat{\vec{X}}^{\prime c})^2+\frac{\theta}{4}(\hat{\vec{P}}^\prime)^2\right)=\omega\left( B_L^{\prime\ddagger} B_L^\prime + B_R^\prime B_R^{\prime\ddagger} +1 \right)~~\textsf{using}~~(\ref{d9})
\end{equation}
We can check that the coefficient of Hamiltonian(\ref{e1}) is indeed $\omega$, by using the values of $\tanh\phi, \alpha$ and $\beta$ so that one gets the spectrum of the same form(\ref{N}) with replacement $\omega_0\rightarrow\omega$:
\begin{equation}
E(j,j_3)=\omega (2j+1)  \label{AA}
\end{equation} 
This demonstrates that $SU(2)$ symmetry is preserved even at a point away from the critical point (\ref{c7}) in the $\mu-\omega$ plane. As we shall show in the next sub-section that this $SU(2)$ symmetry will, however, be broken in the physical oscillator, where unphysical commuting `` position -like" observable $\hat{X_i^c}$ in (\ref{c3}) will be replaced by the physical  non-commuting position operator $\hat{X_i}$ in the Hamiltonian below (\ref{e11}).
\subsection{Physical SHO involving non-commuting coordinates}
Let us finally consider the  Hamiltonian for the physical simple harmonic oscillator involving non-commuting position operators: 
\begin{equation}
\hat{H}_3 = \frac{1}{2\mu}\hat{\vec{P}}^2 + \frac{1}{2} \mu\omega ^2 (\hat{\vec{X}})^2 \label{e11} 
\end{equation}
Using(\ref{f3}), this can be re-expressed as:
\begin{equation}
\hat{H}_3 = \frac{1}{2\mu}\hat{\vec{P}}^2 + \frac{1}{2} \mu\omega ^2 \left[ (\hat{\vec{X}}^c)^2 + \frac{\theta ^2}{4}\hat{\vec{P}}^2 + 2\theta \hat{J_3} \right], \label{e8}
\end{equation}
which can be brought to the same form as $\hat{H}_2$(\ref{c3}), up to a Zeeman term as:
\begin{equation}
\hat{H}_3= \frac{1}{2\mu^\prime}\hat{\vec{P}}^2 + \frac{1}{2} \mu^\prime\omega ^{\prime 2} (\hat{\vec{X}}^c)^2 +\mu\theta\omega ^2\hat{J_3}, \label{1}
\end{equation}
with re-normalized parameters $\mu^\prime$ and $\omega^\prime$, satisfying $\mu \omega ^2=\mu^{\prime} {\omega^{\prime}}^2$, are given by:
\begin{equation}
\frac{1}{\mu^\prime}=\frac{1}{\mu}+\frac{\mu\omega^2\theta^2}{4}~~\textsf{and}~~\omega ^{\prime 2}=\omega^2\left(1+\frac{\mu^2\omega^2\theta^2}{4}\right) \label{e2}
\end{equation} 
The presence of Zeeman term however, breaks its SU(2) symmetry to U(1), as the Hamiltonian does not commute with angular momentum operators $\hat{J}_1$ and $\hat{J}_2$ anymore. But the Hamiltonian is still symmetric under rotation generated by   $\hat{J}_3$. Now we can again do the same canonical transformation(\ref{d7}), with new parameters $\mu'$ and $\omega'$ satisfying 
\begin{equation}
e^{2\phi}=\frac{\mu^\prime\omega^\prime\theta}{2}, \label{A}
\end{equation} 
and keeping in mind that $\hat{J}_3 = \epsilon_{ij}\hat{X}^c_i\hat{P}_j$ remains unaffected under this transformation, to get:
\begin{equation}
\hat{H}_3=\omega^\prime\left( B_L^{\prime\ddagger} B_L^\prime + B_R^\prime B_R^{\prime\ddagger} +1\right) +\dfrac{\mu^\prime \theta {\omega^\prime}^2}{2}\left( B_L^{\prime\ddagger} B_L^\prime - B_R^\prime B_R^{\prime\ddagger} \right) \label{p1}
\end{equation}
The corresponding spectrum can be easily read as
\begin{equation}
E(j,j_3)= \omega^\prime (2j+1) + \theta \mu^\prime{\omega^\prime}^2 j_3 \label{BB}
\end{equation}
The presence of $j_3$-dependent term is clearly due to the presence of the $SU(2)$- symmetry breaking Zeeman term in in (\ref{1}). The occurrence of these renormalised parameters $\omega^\prime$ and $\mu^\prime$ in the spectrum (\ref{BB}) indicates that these parameters, rather than their 'bare' counterparts $\omega$ and $\mu$ occurring in (\ref{e11}), are the observable quantities of the theory. Indeed, it can be easily checked, using that for   $\mu^\prime$ and $\omega^\prime$ at the critical point (\ref{c7}) i.e. $\mu^\prime =\frac{\omega^\prime}{2}=\frac{1}{\sqrt{\theta}}$, the corresponding bare mass  diverges $\mu\rightarrow\infty$ and frequency $\omega\rightarrow 0$. Nevertheless, in order to establish compatibility with \cite{b7}, it is desirable to express (\ref{p1}) in terms of the bare quantities $\mu$ and $\omega$.
Using the definition of $\mu^\prime$ and $\omega^\prime$(\ref{e2}), we can easily write above Hamiltonian (\ref{p1}) as:
\begin{equation}
\hat{H}_3 = \frac{\lambda_+}{2\mu} \left(2B_L^{\prime\ddagger} B_L^\prime + 1\right) + \frac{\lambda_-}{2\mu} \left(2 B_R^\prime B_R^{\prime\ddagger} +1\right), \label{e3}
\end{equation}
where $\lambda_+$ and $\lambda_-$ are given by:
\begin{equation}
\lambda_{\pm}=\frac{1}{2}\left(\mu\omega\sqrt{4+\mu^2\omega^2\theta^2} \pm \mu^2\omega^2\theta \right)
\end{equation}
This clearly reproduces spectrum of \cite{b7} calculated using a different approach. We therefore turn our attention now towards the determination of the ground state. To begin with, let us note that, unlike in the  
commutative case, where the ground $|0\rangle\otimes|0\rangle$ is annihilated by operators $\hat{a}_i$ with $\hat{a}_1$ and $\hat{a}_2$ annihilating the ground states $|0\rangle$ occurring in the left and right sectors respectively in $|0\rangle\otimes|0\rangle$, in the non-commutative case the corresponding ground state $|0\rangle\langle 0|$ is not annihilated by operators $\hat{C}_i$(\ref{e5}) except at the critical point(\ref{c7}). Besides, the form of $\hat{C}_i$ (\ref{f6}) at the critical point (\ref{c7}) clearly indicates that these $\hat{C}_i$'s have simultaneous actions on both the slots. Indeed, the analogous expressions for the commutative case will be combinations $\frac{1}{\sqrt{2}}(\hat{a}_1+\hat{a}_2)$ and $\frac{-i}{\sqrt{2}}(\hat{a}_1-\hat{a}_2)$ respectively. But away from the critical point (\ref{c7}) i.e. for general values of $\mu$ and $\omega$ there must exist another ground state $|0\rangle^\prime\langle 0|^\prime $, related to the older one as in (\ref{e6}), which should be  
annihilated by a transformed $\hat{C}^{\prime}_i$.
\begin{equation}
\hat{C}^{\prime}_i|0\rangle^\prime\langle 0|^\prime=0 ~~; ~~ |0\rangle^\prime\langle 0|^\prime = e^{-\phi(\hat{B}_L^\ddagger \hat{B}_R - \hat{B}_L \hat{B}_R^\ddagger)} |0\rangle\langle 0| \label{x1}
\end{equation}
where $\hat{C}^{\prime}_i$ is defined analogously as in (\ref{f6}) 
\begin{equation}
\hat{C}^{\prime}_1=\frac{1}{\sqrt{2}}\left(\hat{B}^{\prime}_L+{\hat{B}^{\prime\ddagger}}_{R} \right);~~\hat{C}^{\prime}_2=\frac{-i}{\sqrt{2}}\left(\hat{B}^{\prime}_L-{\hat{B}^{\prime\ddagger}}_R\right) \label{e10}
\end{equation}
in terms of $\hat{B}_L ^\prime$ and $\hat{B}_R ^\prime$ introduced in (\ref{d4}). 
Equivalently, 
\begin{equation}
\hat{B}^{\prime}_L |0\rangle^\prime\langle 0|^\prime =0 ~ \text{and} ~ \hat{B}^{\prime\ddagger}_R |0\rangle^\prime\langle 0|^\prime =0 \label{eq82}
\end{equation}
which are clearly not independent of each other, as they are related by Hermitian conjugation. One can easily see at this stage that either of these equations are trivially satisfied by making use of (\ref{z1},\ref{e6},\ref{x1}). But, before going ahead with the straightforward computation of $|0\rangle^\prime\langle 0|^\prime$  and compare with the normalised version of the ground state obtained in \cite{b7}, we would like to point out certain exact parallel between the approach followed here with \cite{b7}. To that end, note that, using these equations (\ref{eq82}) and making use of (\ref{d4}) we get 
\begin{equation}
\hat{b} |0\rangle^\prime\langle 0|^\prime = - \tanh \phi |0\rangle^\prime\langle 0|^\prime \hat{b} 
\end{equation}
and its Hermitian conjugate. Then using (\ref{A}) and the counterpart of (\ref{B}) i.e $\tanh \phi= \frac{1}{\beta^{\prime}} (\alpha^{\prime} - \omega^{\prime})$, where $\alpha^{\prime}$ and $\beta^{\prime}$ defined analogously as in (\ref{C}), with the replacements $\mu \rightarrow \mu^{\prime}$ and $\omega \rightarrow \omega^{\prime}$, yielding
\begin{equation}
(1+\theta\lambda_{+}) \hat{b} |0\rangle^\prime\langle 0|^\prime = |0\rangle^\prime\langle 0|^\prime  \hat{b} \label{eq83}
\end{equation}
just from the  first equation in (\ref{eq82}).
As expected, the second equation of (\ref{eq82}) yields the Hermitian conjugate of (\ref{eq83}) and these pair of equations exactly reproduces the defining equations (\ref{eq60} , \ref{d4}) of ref \cite{b7}. One can then proceed analogously, as in \cite{b7} to get the normalized version of the ground state as 
\begin{equation}
\psi_0 \equiv |0\rangle^\prime\langle 0|^\prime = \frac{1}{\cosh \phi} e^{\gamma(\hat{b}^\dagger \hat{b})} \label{eq89}
\end{equation}
where
\begin{equation}
\gamma = \log(1-\theta\lambda_{-}) = \log(-\tanh \phi)
\end{equation}
so that
\begin{equation}
(\psi_0|\psi_0)=\frac{1}{\cosh^2 \phi} Tr_c(e^{2\gamma \hat{b}^\dagger \hat{b}}) = 1.
\end{equation}
Here, we have made use of the identities 
\begin{equation}
(1+\theta\lambda_{+})(1-\theta\lambda_{-}) = 1 ~ \text{and} ~ \tanh \phi = \frac{\lambda_{+}}{\lambda_{-}} = \frac{1}{1+\theta\lambda_{+}}
\end{equation}
Now expanding (\ref{eq89}), using the complete set of basis $\sum^\infty _{m=0}|m\rangle\langle m| = \mathbb{1}$, as 
\begin{equation}
\psi_0 = \text{sech} \phi \Sigma_{m=0} ^{\infty} (-1)^m (\tanh \phi)^m |m\rangle\langle m| = \text{sech} \phi \left( |0\rangle\langle 0|-\tanh \phi |1\rangle\langle 1| + ... \right) \label{92}
\end{equation} 
we can easily see that the Taylor's expansion of $\tanh \phi $ ,  $\text{sech} \phi$ and their product exactly reproduces (\ref{x1})
\begin{eqnarray}
|0\rangle^\prime\langle 0|^\prime &=&  e^{-\phi (\hat{B}_L^\ddagger \hat{B}_R - \hat{B}_L \hat{B}_R^\ddagger)} |0\rangle\langle 0|  \nonumber \\
&=&(1-\frac{\phi^2}{2!} + \frac{5}{4!}\phi^4 -...)|0\rangle\langle 0| + (\phi-\frac{5\phi^3}{3!} + \frac{5}{4!}\phi^5 -...)|1\rangle\langle 1| + ... =  \psi_0
\end{eqnarray}
Finally, note that the sign alternates from one term to another, because of the presence of the factor $(-1)^m$ in $\psi_0$ in (\ref{92}) and therefore cannot be interpreted as a mixed density matrix, from the perspective of classical Hilbert space $\mathcal{H}_c$ (\ref{i2}).
\section{Time reversal}
We proceed to check how the Hamiltonian $\hat{H}_3$,  defined through (\ref{p1}),  behaves under time reversal. Using the defination of $\hat{X}^{c}_i $
defined through (\ref{f3}), $\hat{X}^{L}_i$ can be written as-
\begin{equation}
\hat{X}^{L}_i = \hat{X}^{c}_i+\frac{\theta}{2}\epsilon_{ij}\hat{P}_j
\end{equation}
Now using the fact that $\hat{X}^{c}_i $, which is mathematically commuting coordinate, does not undergo any change: $\hat{X}^c_i\rightarrow \hat{X}^c_i$  under time reversal and $\hat{P}_i\rightarrow -\hat{P}_i$ 
under time reversal, the following transformation property of $\hat{X}^{L}_i$ under time reversal can be obtained
\begin{equation}
\Theta \hat{X}^{L}_i\Theta^{-1} = \hat{X}^{L}_i+\theta\epsilon_{ij}\hat{P}_j ,
\end{equation}
where $\Theta$ is the time reversal operator. Similarly, we obtain the transformation property of $\hat{X}^{R}_i$ under time reversal to be
\begin{equation}
\Theta \hat{X}^{R}_i\Theta^{-1} = \hat{X}^{R}_i-\theta\epsilon_{ij}\hat{P}_j
\end{equation}
Further, we proceed to check the transformation property of $\hat{B}_{L/R}$ (defined in section II)
\begin{eqnarray}
\Theta \hat{B}_{L/R}\Theta^{-1} = \hat{B}^{\ddagger}_{R/L}\\
\Theta \hat{B}^{\ddagger}_{L/R}\Theta^{-1} = \hat{B}_{R/L}
\end{eqnarray}
So that the Hamiltonian $\hat{H}_3$ (\ref{e11}) and angular momentum $\hat{J}_3$, under time reversal transform to
\begin{eqnarray}
\Theta \hat{H}_{3}\Theta^{-1} &=& \omega'(\hat{B}'^{\ddagger}_L\hat{B}'_L + \hat{B}'_R\hat{B}'^{\ddagger}_R+1)-\frac{\mu'\theta\omega'^2}{2}(\hat{B}'^{\ddagger}_L\hat{B}'_L - \hat{B}'_R\hat{B}'^{\ddagger}_R)\neq \hat{H}_3 \\
\Theta \hat{J}_{3}\Theta^{-1} &=& -(\hat{B}'^{\ddagger}_L\hat{B}'_L - \hat{B}'_R\hat{B}'^{\ddagger}_R) = -\hat{J}_{3}
\end{eqnarray}
This clearly indicates that time reversal symmetry is broken and that the Zeeman term is solely responsible for breaking the both SU(2) and time reversal symmetry.

\section{Conclusion}
 The tensor product structure of the quantum Hilbert space $\mathcal{H}_q$ as $\mathcal{H}_{c}\otimes\tilde{\mathcal{H}}_{c}$ is exploited to construct Schwinger angular momentum operators, under which the commuting position like operator $\hat{X}^{c}_{i}$'s are shown to transform covariantly as happens in 3-d case \cite{b13}. A non-physical planar harmonic oscillator is constructed out of the above mentioned commuting position like operator. The ground state of this harmonic oscillator is then shown to coincide with the ground state $|0\rangle\langle0|$ of $\mathcal{H}_q$ only for a particular choice of the parameters mass ($\mu$) and angular frequency ($\omega$) which we refer to be critical point in $\mu-\omega$ plane. Furthermore, the Hamiltonian is shown to be manifestly SU(2) invariant only at this critical point. For the parameters value away from the critical point a canonical transformation (scaling) is required. This is shown to arise naturally by  carrying out a Bogoliubov 
transformation. The 
corresponding ground state is then obtained by a dilatation operator implemented through a unitary transformation. Finally, the physical Hamiltonian involving non-commutative position operators is constructed. It is shown to be of the same form as that of the unphysical oscillator mentioned above, except that a new Zeeman term involving $\hat{J}_3$ pops up. With this both mass and angular frequency parameter undergoes renormalisation, which are eventually identified with physical mass and angular frequency. The presence of Zeeman term is therefore shown to be solely responsible for violating both SU(2) and time reversal symmetry \cite{b7}.
 
\section*{Acknowledgments}
KK thanks three Indian Academy of Sciences for financial support to carry out summer project at the S N Bose Centre Kolkata.
SP thanks the Department of Science and Technology (DST), Government of India, for providing financial support through INSPIRE fellowship. Both of them thanks the  authorities of S N Bose Centre for their kind hospitality during the course of this work. Finally, one of us, BC thanks Prof. F. G. Scholtz for useful comments.

\end{document}